\documentstyle[aps,multicol,amssymb,epsfig]{revtex}

\begin{document}
\title{
Requirement of optical coherence for \\
continuous--variable quantum teleportation
}

\author{Terry Rudolph$^{1}$\cite{address} and Barry C.\ Sanders$^{1,2}$}

\address{
        $^1$The Erwin Schr\"odinger International Institute
                for Mathematical Physics, Boltzmanngasse 9,
                1090 Vienna, Austria    \\
        $^2$Department of Physics,
                Macquarie University, Sydney, New South Wales 2109, Australia}

\date{\today}

\maketitle

\begin{abstract}
We show that the sender (Alice) and the receiver (Bob) each require
coherent devices in order to achieve unconditional continuous variable
quantum teleportation (CVQT), and this requirement cannot be achieved
with conventional laser sources, linear optics, ideal photon detectors,
and perfect Fock state sources. The appearance of successful CVQT in
recent experiments is due to interpreting the measurement record
fallaciously in terms of one preferred ensemble (or decomposition) of
the correct density matrix describing the state. Our analysis is
unrelated to technical problems such as laser phase drift or finite
squeezing bandwidth.
\end{abstract}

\begin{multicols}{2}

Quantum teleportation, first proposed as a method for teleporting an unknown
spin--$1/2$ quantum state from a sender, ``Alice'', to a receiver,
``Bob''\cite{Ben93}, has been extended to continuous--variable quantum
teleportation (CVQT)\cite{Vai94,Bra98,Fur98,Bra00}.  In this variant of quantum
teleportation, Alice and Bob share a nonlocal entangled two--mode field which,
ideally, should have perfect correlations between in--phase quadrature
(`position'~$x$) and out--of--phase quadrature (`momentum'~$p$), although the
requirement of perfect correlations can be relaxed somewhat\cite{Bra98}.
Unconditional teleportation of coherent states was recently claimed to have
been experimentally demonstrated\cite{Fur98}, with two--mode squeezed light
providing the nonlocal entangled resource shared between Alice and Bob.
Although experimental advances towards CVQT have been achieved, we show that
genuine CVQT cannot be achieved using conventional laser sources, due to an
absence of intrinsic optical coherence\cite{Mol97}. This incoherence is not due
to phase drift but is rather a consequence of entanglement between the laser
medium and the output field. Herein we exclusively use the term coherence to
refer to coherent superpositions of Fock states. In this letter, we establish
that optical coherence is \emph{required} for true CVQT to succeed, that
conventional lasers do not meet this criterion, and therefore that
unconditional CVQT is not possible with conventional lasers.

Clear criteria for evaluating the success of a teleportation experiment have
been established\cite{Bra00}, and we emphasize three relevant criteria for
determining whether CVQT is possible at all with conventional laser sources:
(a)~the states to be teleported should be unknown to Alice and Bob and supplied
by an actual third party Victor; (b)~Alice and Bob share {\it only} a nonlocal
entangled resource and a classical channel through which Alice transmits her
measurement results to Bob; and (c)~entanglement should be a verifiable
resource.
In particular we will show that the experiment of Ref.~\cite{Fur98} does not
satisfy any of these criteria and establish that any scheme based on
conventional laser sources cannot simultaneously satisfy all three criteria.

\begin{figure}
\centerline{\epsfig{figure=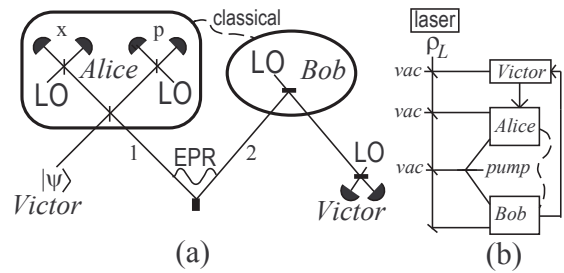,width=100mm}}
\caption{(a)~Ideal CVQT.  Alice mixes, at a 50/50 beam splitter,
Victor's state with her portion (component~1) of a two--mode squeezed vacuum
state that is shared with Bob (component~2). In the ideal limit, the two--mode
squeezed field is a pure entangled state with a Wigner function proportional to
$\delta(x_1+x_2)\delta(p_1-p_2)$, for $x_i$ ($p_i$) the in--phase
(out--of--phase) quadrature of field component~$i\in\{1,2\}$, thereby acting
as a true EPR state.  Alice mixes the two beamsplitter output fields  with
local oscillators (LOs) assumed to be in (infinitely strong) coherent states,
thereby yielding true $(x,p)$ quadrature--phase homodyne measurements. She
transmits outcome~$(x,p)$ to Bob via a classical channel, which he uses to
transform component~2 of the EPR state by mixing with a LO at a beam splitter.
(b)~An experimental simulation of CVQT using an incoherent (laser) source such
as that adopted in Ref.~[4].  The laser field is shared by all parties.}
\label{scheme}
\end{figure}

The ideal scheme for CVQT\cite{Bra98} is explained in
Fig.~\ref{scheme}(a), and a schematic and explanation of the
experimental simulation of CVQT is provided in Fig.~\ref{scheme}(b). We
emphasize that the  scheme of Fig.~\ref{scheme}(a) presupposes the
availability of large amplitude coherent states to serve as local
oscillators (LOs) for both Alice and Bob, and as a pump to produce the
two-mode squeezed state. One should also assume that Victor has the
ability to produce truly coherent states. In the experimental
simulation of CVQT~\cite{Fur98}, the same laser source is used for
(i)~producing Victor's state for teleportation, (ii)~pumping the
nonlinear crystal that produces a two--mode squeezed light field
serving as the shared Einstein-Podolsky-Rosen (EPR) state,
(iii)~supplying the LO
fields for both of Alice's homodyne measurements, and (iv)~providing
Bob with a coherent field to mix with his portion of the EPR beam to
reconstruct Victor's state, as depicted in Fig.~\ref{scheme}(b).

In all discussions relating to optical CVQT, lasers have been presumed
sources of (Gaussian) coherent states \cite{Gla63} $\vert\alpha
e^{i\phi} \rangle = \exp(-\alpha^2/2) \sum_{n=0}^\infty (\alpha^n
e^{in\phi}/\sqrt{n!}) \vert n \rangle$. In quantum optics it is, in
fact, standard to assume that the output of an ideal single--mode laser
operating well above threshold is given by a coherent state. However,
it is also well known that the phase $\phi$ of this laser field is
completely unknown (not to be confused with the physically
irrelevant global phase of a wavefunction).  As such, the state of the
laser output field~$\rho_L$ \emph{must} be represented by a density
matrix with phase eliminated by integrating over the distribution
representing this lack of classical information. The resulting density
matrix,
\begin{mathletters}
\label{laserstate}
\begin{eqnarray}
\label{laserstatea}
\rho_L &=& \int_0^{2\pi} \frac{d\phi}{2\pi} \, \vert \alpha e^{i\phi} \rangle
                \langle \alpha e^{i\phi} \vert~
                      \\
        &=& e^{-\alpha^2} \sum_{n=0}^\infty
 \frac{\alpha^{2n}}
        {n!} \vert n \rangle \langle n \vert~,
\end{eqnarray}
\end{mathletters}
is the correct quantum mechanical description of an ideal laser mode output.
As $\rho_L$ is diagonal in the energy eigenstate (Fock) basis it is, in our
precise terminology, not coherent.

It is common in quantum optics to assume that the single--mode laser field is
in a coherent state $\vert \alpha e^{i\phi} \rangle$ to a good approximation.
From Eq.~(\ref{laserstate}a) it is clear why this works so well: this
particular decomposition (or ensemble) of the reduced density matrix for the
laser light corresponds to a mixture of coherent states with equally likely
phase, and the success or failure of experiments in quantum optics is never
dependent on knowing the {\em presumed} initial phase of the laser. More
precisely, conventional measurements on optical fields, such as homodyne
detection using lasers, involve mixing of different incoherent fields and
subsequent detection by energy absorption in photodetectors: all such
measurements are completely insensitive to any optical coherences\cite{Spe01}.

We have mentioned the entanglement between the laser field and the state of the
lasing medium as the source of the laser field phase indeterminacy. In
\cite{Mol97}  M\o lmer explains carefully the absence of optical coherence for
conventional lasers and produces a convincing analysis that shows optical
coherence is a ``convenient fiction'' in quantum optics experiments. His
analysis proceeds as follows. The gain medium of a laser consists of
distinguishable dipoles (e.g. atoms), which are initially in a thermal
distribution of their energy eigenstates and, therefore, initially possess a
mean dipole moment of zero. This gain medium is incoherently pumped (the
pumping field too is in a classical distribution of energy eigenstates and
therefore has a zero mean electric field), and the standard interaction between
the atoms and this field produces a large {\em entangled} state of the laser
mode, pumping modes, gain medium and so on (with of course the build up of a
large number of photons in the laser mode by stimulated emission). However, we
are generally interested only in the light in the laser mode, and the correct
description of this field state is via the reduced density matrix obtained by
tracing out all other degrees of freedom. All processes involved are energy
conserving, and the initial states of all participating physical systems (gain
medium, pump modes, etc.) begin with no coherence: they are initially described
by diagonal density matrices in their energy eigenstate bases. Evidently,
tracing out the gain medium, pump modes, etc.\ of the large entangled state
produced under these conditions {\it cannot} leave the laser mode in a state
with any non-zero off-diagonal elements in its density matrix, let alone in a
pure coherent state.  More concisely, one cannot obtain states with quantum
uncertainty in energy (coherent superpositions of Fock states) via physical
processes which have only classical uncertainty in energy (that is, are
diagonal in the Fock basis).

The intrinsic indeterminacy of the laser phase is unrelated to the technical
issue of phase drift: phase stabilization methods via mixing of laser beams do
\emph{not} establish the existence of a phase for the laser. The importance of
controlling phase drift is that \emph{relative} phases of laser beams,
generally (but not exclusively) spawned from the same source as in
Fig.~\ref{scheme}(b), have reduced drifts. This amounts to stabilization of the
mode under consideration and is important because quantum optical interference
experiments rely on \emph{indistinguishability} (not coherence) for their
success. Stabilizing this phase drift cannot introduce coherences into the
overall density matrix.

The decomposition (\ref{laserstate}a) to coherent states notwithstanding,
selecting one coherent state as the de facto state of the laser is clearly a
commission of the ``partition ensemble fallacy'' (PEF) \cite{Kok00}. Within the
framework of standard quantum mechanics, the density matrix is the complete
description of the quantum state \cite{Per98}, and there is no reason to accord
preferential treatment to one particular decomposition of the infinite number
of equivalent decompositions for any mixed state. Avoiding preferential
decompositions is clearly important for quantum information processing, which
should be formulated operationally. We emphasize that the interpretation of
experimental data under the assumption that the ``real'' state of the laser is
actually one element of one decomposition of $\rho_L$ yields a fallacious
interpretation of the measurement record in terms of ``what really happened''.

Tomographic reconstruction of Gaussian states with well--defined
phase\cite{Smi93} provides a simple example of how a failure to formulate
operationally can lead to incorrect retrodiction of results. Such experiments
generally  rely on autocorrelation measurements in which the reconstructed
state and the LO used for the homodyne measurements are derived from the same
laser source, similar to the CVQT case considered in Fig~\ref{scheme}(b). As
the a priori assumption is that the laser produces a coherent state, the
experimental data is interpreted in this context, and the tomographically
reconstructed state is, therefore, interpreted as a coherent state.  Since
decomposition~(\ref{laserstate}a) of $\rho_L$ allows one to describe the
initial laser state as a coherent state, and the experiment relies only on
phase differences and not the overall initial phase of the laser, clearly no
contradiction with this partitioning assumption can be achieved.  It is
convenient to assume that the laser field is in a coherent state, but such
tomographic reconstruction does not ``prove'' this.

We are of course \emph{not} asserting that production of coherent states of
light is impossible: basic quantum electrodynamics shows that a classical
oscillating current can produce coherent states\cite{Gla63}. As a classical
current can be fully measured without disturbance, it provides the classical
information necessary to specify a unique coherent state. Production of true
coherent states is also possible if coherent measurements can be performed on
one component of an entangled system. For example, a measurement on a two-level
atom (with energy eigenstates $|g\rangle,|e\rangle$), coupled to a single
photon cavity mode (with energy eigenstates $|0\rangle,|1\rangle$) in the
entangled state $(|g,1\rangle+|e,0\rangle)/\sqrt{2}$, can be used to project
the photon into a state with coherence, e.g. $(|0\rangle+|1\rangle)/\sqrt{2}$.
However, such a measurement cannot be in the energy eigenstate basis of the
atom, it necessarily requires coherence. Similarly, a coherent measurement on
the gain medium of the laser, while technically challenging, could in principle
project the laser output into a coherent state. However, processing the output
of the conventional laser via linear optics, ideal photon detectors and perfect
Fock state photon sources
\emph{cannot} be used to project the laser mode into a coherent state.
CVQT as envisioned in all
optical proposals thus far has been teleportation of phase and amplitude
information of a coherent field. We therefore assert that genuine CVQT requires
{\it coherent devices}, that is, devices capable of generating true coherence,
and these are not a feature of current CVQT experiments.


We address the three criteria for successful CVQT, beginning with
criterion~(a). Splitting a coherent state $\vert\alpha e^{i\phi}\rangle$ at a
lossless beam splitter with reflectivity~$r$ and transmissivity~$t$ yields a
product state $|ir\alpha e^{i\phi}\rangle\otimes|t\alpha e^{i\phi}\rangle$, and
the two output modes are, therefore, independent; one mode could be used by
Victor. However, if the input to the beamsplitter is $\rho_L$, the resultant
state is {\it not} of the form $\rho_1
\otimes \rho_2$ but rather
\begin{eqnarray}
\label{beamsplitter}
\rho_{\rm BS}
    &=&\int_0^{2\pi} \frac{d\phi}{2\pi} |ir\alpha e^{i\phi}\rangle
    \langle ir\alpha e^{i\phi}\vert
    \otimes |t\alpha e^{i\phi}\rangle\langle t\alpha e^{i\phi}|
        \nonumber \\
    &=& e^{-\alpha^2}\sum_{m,n=0}^\infty
    i^{n-m} \frac{(r\alpha)^{n+m}}{\sqrt{n!m!}}
    \sum_{r,s=0}^\infty \frac{(t\alpha)^{r+s}}{\sqrt{r!s!}}
        \nonumber \\    &&  \times
    \delta_{m+r,n+s}\vert m \rangle \langle n \vert
    \otimes \vert r\rangle\langle s\vert~,
\end{eqnarray}
with significant correlation between the two output modes. This clearly
violates criterion~(a). Decomposing state~(\ref{beamsplitter}) to a specific
product coherent state, with one of these two coherent states to be teleported,
commits, once again, the PEF\cite{Kok00}.

Criterion~(b) stipulates that Alice and Bob should share {\it only} a nonlocal
entangled resource and a classical communication channel.  As the laser phase
is indeterminate (more precisely it is a property  whose ``existence'' depends
on ascribing reality to one ensemble of $\rho_L$), Alice
cannot include information on the phase of her laser source in the classical
information she transmits to Bob,
using linear optics, ideal photon detectors, and perfect Fock state sources.
In the existing experiment Alice shares with
Bob a laser beam split from the same source. He uses this beam to apply a
unitary displacement operation (determined by Alice's transmitted classical
information) to reproduce Victor's state. Without sharing this beam (or
something similar but equally incoherent such as two slave beams locked to one
master laser), Alice's and Bob's operations are completely uncorrelated. We
maintain therefore that the absence of intrinsic optical coherence in a
conventional laser necessitates the existence of this manifestly quantum
channel. A classical channel cannot replace this because, although from the
first decomposition of Eq.~(\ref{beamsplitter}) $\rho_{BS}$ is separable, it is
locally preparable
\emph{only if} Alice and Bob each have independent coherent sources. To see
this, note that if Alice and Bob each have only non--coherent devices (as
defined earlier), classical communication only permits them to prepare mixed
states (in the Fock basis), of the form $\rho=\sum_{ij} c_{ij} |i\rangle\langle
i |\otimes |j\rangle\langle j\vert\neq\rho_{BS}$. In the experiment of
Ref.~\cite{Fur98}, the majority of photons in the teleported state that Bob
supplies to Victor (for verification) actually arrive from this extra quantum
state shared between Alice and Bob. Without this shared field, the ``convenient
fiction'' of coherence cannot be maintained. This quantum channel is in fact
necessarily capable of transmitting the state Victor provided to Alice on its
own; Victor would certainly be entitled to be suspicious of such a scheme.
Teleportation as envisioned in
\cite{Ben93} clearly requires that the only shared quantum state is the
entangled state and Bob applying a unitary operation with only a classical
device. In effect, without truly coherent devices the analogue of the single
qubit rotations in\cite{Ben93} cannot be performed on arbitrary single mode
states of light.

We now consider criterion~(c).  Conventionally in quantum optics, it is assumed
that the nonlinear crystals used to produce squeezed states are pumped by a
strong, pure coherent state. This results in a two--mode squeezed field,
\begin{equation}
\label{squeezed:purestate}
\vert \eta e^{i\phi} \rangle
    = \sqrt{ 1 -  \eta^2 }
        \sum_{n=0}^\infty \eta^n e^{in\phi}\vert n \, n \rangle~,
\end{equation}
with $\phi$ the phase of the pump field. However, the pump field actually
originates from~$\rho_L$ of Eq.~(\ref{laserstate}); hence, the two--mode output
state from the nonlinear crystal is given by
\begin{equation}
\label{squeezed:state}
\rho_S
        = \int_0^{2\pi} \frac{d\phi}{2\pi} \, \vert \eta e^{i\phi} \rangle
                \langle \eta e^{i\phi} \vert
                \nonumber       \\
    = \left( 1 - \eta^2 \right) \sum_{n=0}^\infty
        \eta^{2n} \vert n \, n \rangle \langle n \, n \vert
\end{equation}
which is clearly diagonal in the Fock state basis. It is therefore separable;
Alice and Bob can prepare the state locally and consequently violate
criterion~(c). That $\rho_S$ is diagonal arises because the pump field is
diagonal in the Fock basis (i.e., $\rho_P = \sum \varrho_n \vert n
\rangle_P \langle n \vert $), and thus has a classical energy spread
or uncertainty. A nonlinear crystal pumped by a state with no energy
uncertainty (i.e.~$\rho_P=\vert N \rangle _P\langle N\vert $) transforms the
pure state input $|N\rangle_P|0 \, 0\rangle$, with the two EPR fields in a
vacuum state, into a superposition of states $ \{ \vert N-m\rangle_P \vert m \,
m\rangle \}$. Tracing over the pump field state~$\rho_P$ destroys coherences in
the reduced density matrix for the  two squeezed modes, and results in the
separable state (\ref{squeezed:state}). A two--mode squeezed state may
\emph{appear} to be pure,  for example in tomographic reconstructions using
balanced homodyne detection\cite{Ray96}. However, such reconstruction of a
two--mode Gaussian (squeezed) state is another example of an invalid
retrodiction by not properly applying the operational formulation, as discussed
earlier.  Thus, the pure state representation of the two--mode field as an
entangled state is, in M\o lmer's terminology, ``a convenient fiction''.

 While we have focused here on homodyne type experiments such
as\cite{Fur98}, Molmer's argument applies equally well to systems of
independent lasers interacting through conventional quantum optics devices.
Such experiments cannot create nor demonstrate single mode coherence
\cite{Mol97}. We assert, therefore, that genuine CVQT will require either new
proposals not based on an assumption of coherence or new sources of optical
radiation.

We conclude by describing how aspects of the experiment of \cite{Fur98} would
be viewed if, instead of ascribing reality to the first decomposition
(\ref{laserstate}a) of $\rho_L$, we equally fallaciously ascribe reality to
the second decomposition (\ref{laserstate}b). We would then assert that the
laser is in a pure number state $|N\rangle$ (although we are unsure
exactly which one). When a pure number state encounters a beam splitter it
produces an entangled state. Thus we would conclude that Victor's state is
entangled with both Alice's and Bob's LOs. These in turn are
entangled with the downconversion pump beam (as described above), and this is
entangled with the two squeezed modes. The experiment could now be claimed as a
demonstration of an assortment of entanglement effects, although certainly not
CVQT. Since the two decompositions of (\ref{laserstate}) describe exactly the
same density matrix this description cannot be contradicted by the measurement
record. Our view of course is that both this and the claimed interpretation
are equally incorrect.

In summary, we have established that unconditional CVQT as claimed in
Ref.~\cite{Fur98} is not possible using conventional lasers because
these are not genuine coherent sources. CVQT as envisioned in all
optical proposals thus far has been teleportation of phase and
amplitude information of a coherent field.  The requirement of true
optical coherence is our central result. We have shown that the
experimental simulation of CVQT in \cite{Fur98} cannot be readily
considered ``unconditional teleportation'' without committing the
partition ensemble fallacy (PEF). Three criteria for claiming
unconditional CVQT\cite{Bra00} have been shown to be violated for
experiments using conventional laser sources.

We acknowledge useful discussions with S.\ D.\ Bartlett, C.\ Simon, J.\ Sipe,
R.\ Spekkens, and T.\ Tyc.  This research has been partially supported by an
Australian Research Council Large Grant.
\end{multicols}

\end{document}